\documentclass[]{aastex6}

\usepackage{multirow}
\fullcollaborationName{The Friends of AASTeX Collaboration}

\begin{document}
\fontsize{10pt}{10pt}{Comments: Accepted for publication in The Astrophysical Journal. 10 pages, 8 figures, 1 tables.}
\title{Fermi-LAT detection of A new starburst galaxy candidate: IRAS 13052-5711}

\author{Yunchuan Xiang\altaffilmark{1}\altaffilmark{2}, Qingquan Jiang\altaffilmark{1} and Xiaofei Lan\altaffilmark{1}}

\altaffiltext{1}{School of Physics and Astronomy, China West Normal University, Nanchong 637009, People's Republic of China;\qquad \qquad  \qquad \qquad \qquad \qquad
\quad qqjiangphys@yeah.net;\quad lan-x-f@163.com;\quad xiang{\_}yunchuan@yeah.net}
\altaffiltext{2}{Department of Astronomy, Yunnan University, and Key Laboratory of Astroparticle Physics of Yunnan Province, Kunming, 650091, China}



\begin{abstract}
A likely starburst galaxy (SBG), IRAS 13052-5711, which is the most distant SBG candidate discovered to date, was found by analyzing 14.4 years of data from the Fermi large-area telescope (Fermi-LAT). This SBG's significance level is approximately 6.55$\sigma$ in the 0.1-500 GeV band.
Its spatial position is close to that of 4FGL J1308.9-5730, determined from the Fermi large telescope fourth-source Catalog (4FGL).
Its power-law spectral index is approximately 2.1, and its light curve (LC) for 14.4 years has no significant variability. 
These characteristics are highly similar to those of SBGs found in the past.
We calculate the SBG's star formation rate (SFR) to be 29.38 $\rm M_{\odot}\ yr^{-1}$, which is within the SFR range of SBGs found to date. 
Therefore, IRAS 13052-5711 is considered to be a likely SBG. In addition, its 0.1-500 GeV luminosity is (3.28 $\pm$ 0.67) $\times 10^{42}\ \rm erg\ s^{-1}$, which deviates from the empirical relationship of the $\gamma$-ray luminosity and the total infrared luminosity. 
 We considered a hadronic model to explain the GeV  spectrum of IRAS 13052-5711.
\end{abstract}  

\section{Introduction} \label{sec:intro}
The active galactic nucleus (AGN) is considered to be the most common object with high-energy $\gamma$ rays outside the Milky Way.
 In addition, star-forming galaxies (SFGs) are also a type of famous extragalactic source detected to have high-energy radiation \citep{Ackermann2012}. 
Thus far, SFGs observed in the GeV band include LMC \citep{Abdo2010}, M31, SMC \citep{Nolan2012}, and M33 \citep{Xi2020a}. 
Starburst galaxies (SBGs) are a type of source with a higher SFR than SFGs \citep{Ackermann2012}. 
The currently observed SBG candidates in the high-energy band include NGC 253, M82 \citep{Abdo2010},  NGC 4945, NGC 1068 \citep{Ackermann2012}, NGC 2146 \citep{Tang2014}, Arp 220 \citep{Peng2016}, Circinus \citep{Hayashida2013}, NGC 3424, UGC 11041 \citep{Peng2019}, Arp 299 \citep{Xi2020a}, NGC 2403 \citep{Ajello2020}. 
SBG candidates currently observed in the high-energy band are NGC 253, M82 \citep{Abdo2010},  NGC 4945, NGC 1068 \citep{Ackermann2012}, NGC 2146 \citep{Tang2014}, Arp 220 \citep{Peng2016}, Circinus \citep{Hayashida2013}, NGC 3424, UGC 11041 \citep{Peng2019}, Arp 299 \citep{Xi2020a}, NGC 2403 \citep{Ajello2020}. 
Among them, Circinus, UGC 11041, and NGC 3424, as controversial objects,  have been excluded from the recent statistical studies \citep{Xi2020b, Ajello2020, Xiang2021b}, 
mainly because their LCs exhibit various degrees of variability and their GeV luminosities deviate  considerably from two empirical relationships: $L_{\rm \gamma}$-$L_{\rm 8-1000\ \mu m}$ \citep{Thompson2007} and $L_{\rm \gamma}$- $L_{\rm 1.4\ GHz}$ \citep{Yun2001}; their $\gamma$-ray radiations are currently considered to be dominated by the AGN  \citep{Hayashida2013, Guo2019, Peng2019, Ajello2020}.
In addition, the spatial location of NGC 2403 is close to that of a supernova, 
SN 2004dj, and the origin of $\gamma$ rays at this location is also controversial \citep{Xi2020b, Ajello2020}.

As a type of high-energy celestial body with a high SFR, SBGs are regarded as important probes for studying the evolution of stars outside the Milky Way; they also provide 
 an excellent extreme environment for studying the origin, acceleration, and evolution 
process of cosmic-ray particles (CRs) inside SBGs \citep{Pohl1993,Pohl1994,Wang2018,Peretti2019,Xiang2021b}.
According to current research, the origin of high-energy radiation from SBGs has two possible explanations.
One is linked to the star formation and invokes hadronic interactions of cosmic rays with the ambient gas, whose acceleration is due to SNR \citep{Pohl1993, Pohl1994, Lacki2011, Abramowski2012}; the other is the contribution of high-energy radiation from the internal AGN by continuous accretion from the central massive black hole, in addition to the former’s contribution, for example,
 Arp 299 \citep{Xi2020a}, NGC 3424 \citep{Peng2019}, Circinus \citep{Hayashida2013, Guo2019}.

IRAS 13052-5711, a luminous infrared galaxy, was originally discovered by \citet{Helou1988}
and was recorded in the Infrared Astronomical Satellite (IRAS) Catalog, which provides the average flux density of this object from the bands of 12, 25, 60, and 100 microns. 
\citet{Strauss1992} reported its redshift to be 0.02, based on the observation of IRAS, and its distance as 93.80 Mpc \citep{Crook2007}. \citet{Sanders2003} calculated its 8-1000 $\mu$m luminosity ($L_{\rm 8-1000\ \mu m}$) to be 10$^{11.34}L_{\odot}\sim 8.38 \times 10^{44}$ $\rm erg\ s^{-1}$. 
To date, no detailed studies have reported the GeV emissions of IRAS 13052-5711.
In a previous study, based on the analysis result of 14.4  years of  Fermi-LAT data, 
we found that significant high-energy $\gamma$-ray radiation is present at the position of IRAS 13052-5711, which is only 0$^{\circ}$.094 away from 4FGL J1308.9-5730. 
The current limited angular resolution of Fermi-LAT is approximately 5$^{\circ}$.0 at 100 MeV and 0$^{\circ}$.1 above 20 GeV \citep{Abdollahi2020}. The small deviation in the distance between IRAS 13052-5711 and 4FGL J1308.9-5730 and the exact origin of the $\gamma$-ray radiation in the location of IRAS 13052-5711 have not yet been determined. 
Therefore, this provides us with a strong motivation to explore the origin of the $\gamma$-ray emission in the direction of IRAS 13052-5711. The second section of this paper presents the data analysis, the third section presents a discussion on the results, and the fourth section summarizes the work.

\section{\rm  Fermi-LAT Data Analysis}
\subsection{\rm Data Reduction}
For this analysis, the region of interest (ROI) was a 20$^{\circ}$.0 $\times$ 20$^{\circ}$.0 square centered at (R.A. = 197$^{\circ}$.078, Decl. = -57$^{\circ}$.458; labeled as 4FGL J1308.9-5730 in 4FGL).  
The energy range of the photon events was 100 MeV to 500 GeV; 
the observation period from 2008-08-04 to 2023-01-12 was selected to perform this analysis. 
 Moreover, we selected the Pass 8 data with the ``Source'' event class (evtype = 3 \& evclass = 128) and excluded events with zenith angles above 90$^{\circ}$ to minimize the contamination from the Earth Limb. 
Version v11r5p3 package\footnote{https://fermi.gsfc.nasa.gov/ssc/data/analysis/software/}from Fermi Science Tools (Fermitools) was selected  for this analysis. For the instrumental response function, we adopted ``P8R3{\_}SOURCE{\_}V3''. 
The data analysis thread provided by the Fermi Science Support Center\footnote{http://fermi.gsfc.nasa.gov/ssc/data/analysis/scitools/} was strictly followed. 
All the data were divided into 37 logarithmic energy bins, and each spatial pixel size was set to be 0$^{\circ}$.1 $\times$ 0$^{\circ}$.1. 
Using the binned maximum likelihood method, we selected all known sources of the ROI listed in 4FGL and the two diffuse background templates, 
which are the galactic diffuse emission ({\tt gll{\_}iem{\_}v07.fits}) and the isotropic extragalactic emission ({\tt iso{\_}P8R3{\_}SOURCE{\_}V3{\_}v1.txt})\footnote{http://fermi.gsfc.nasa.gov/ssc/data/access/lat/BackgroundModels.html}, for this analysis  \citep{Abdollahi2020}.
 All spectral parameters of sources within 5$^{\circ}$.0 from the center of the ROI and the normalizations of the  two diffuse backgrounds were left free. 

\subsection{\rm Cross-identification of spatial locations} \label{sec:data}

\begin{figure}
  \includegraphics[width=60mm,height=60mm]{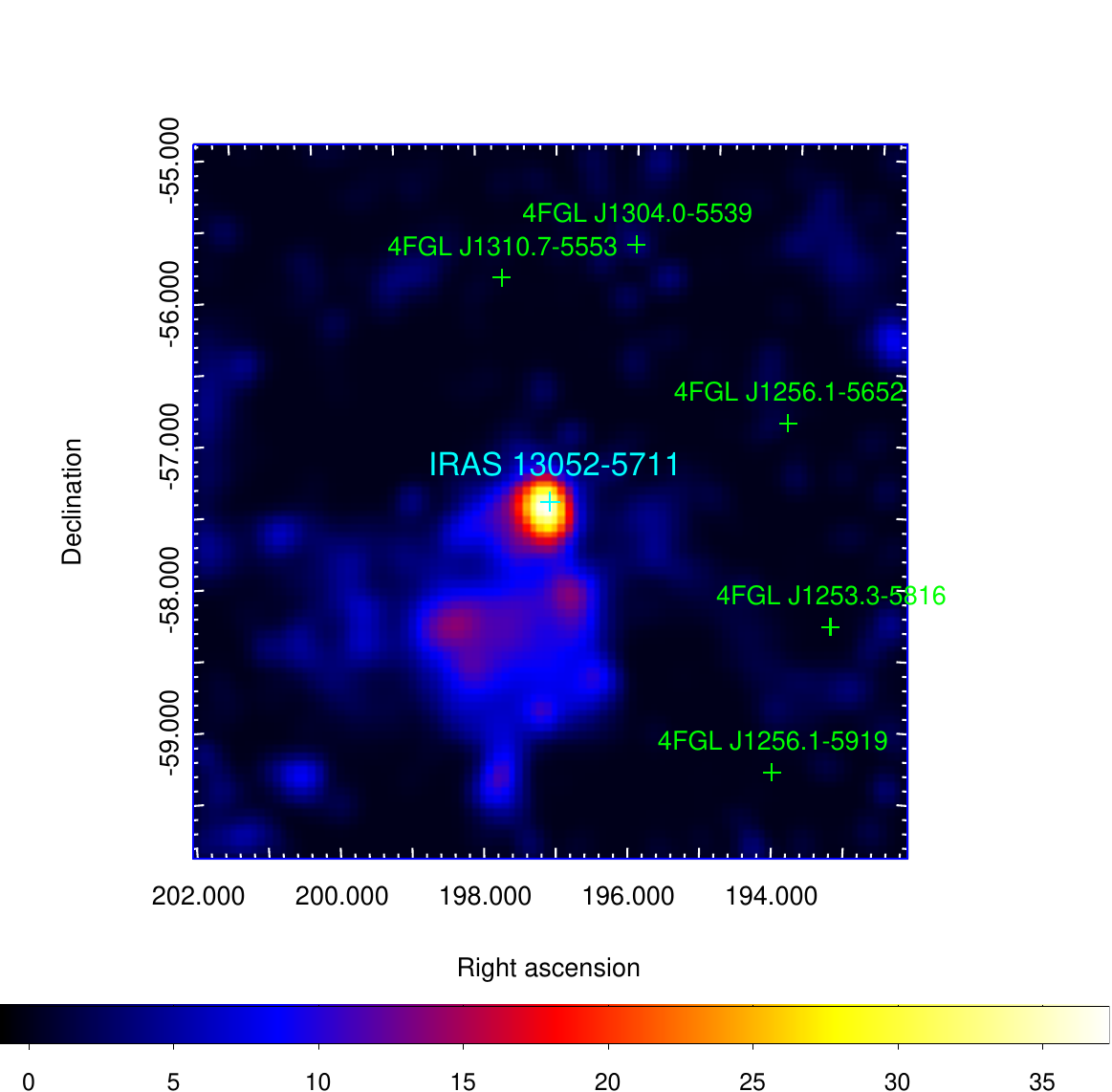}
  \includegraphics[width=60mm,height=60mm]{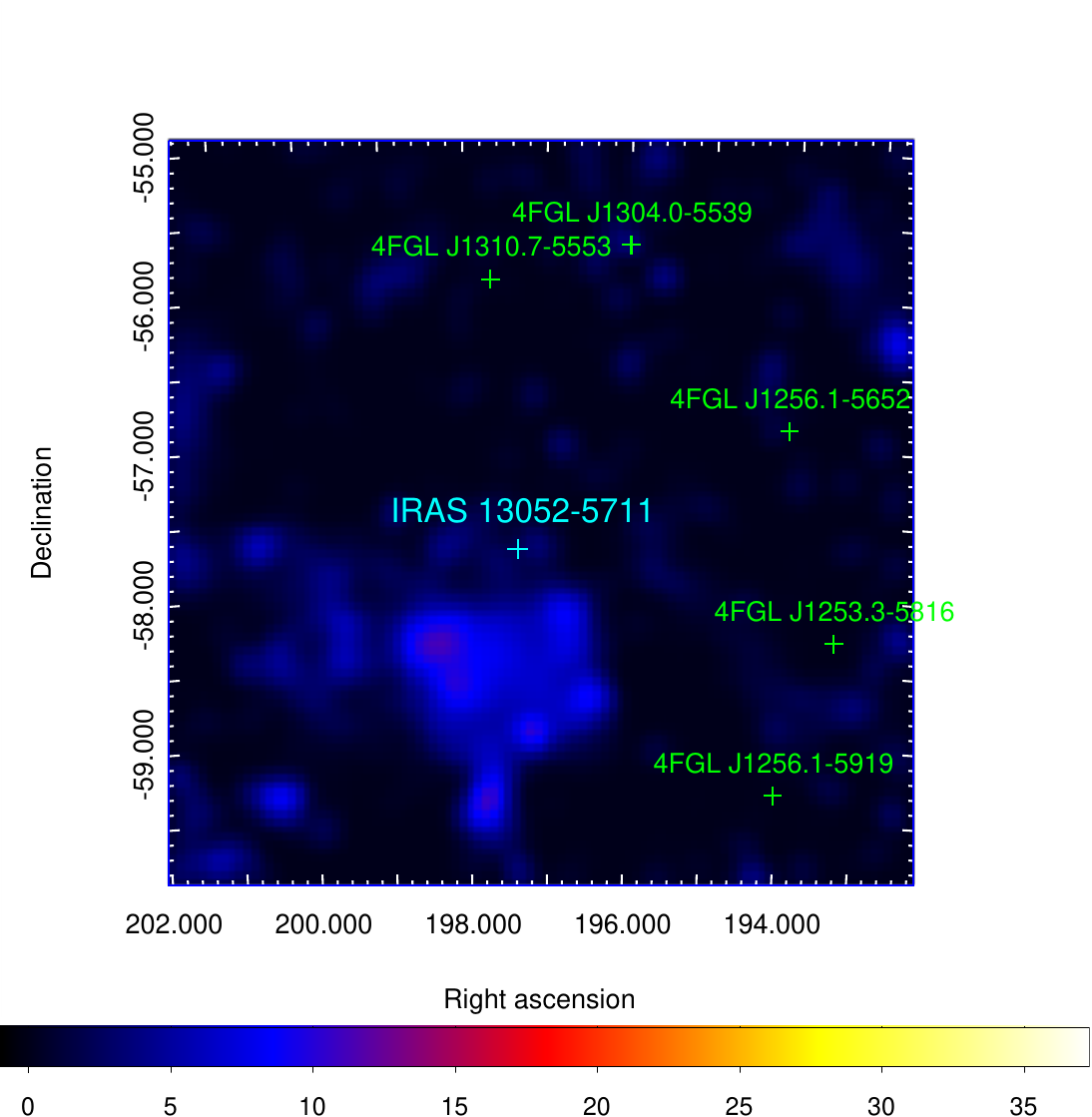}
  \includegraphics[width=60mm,height=60mm]{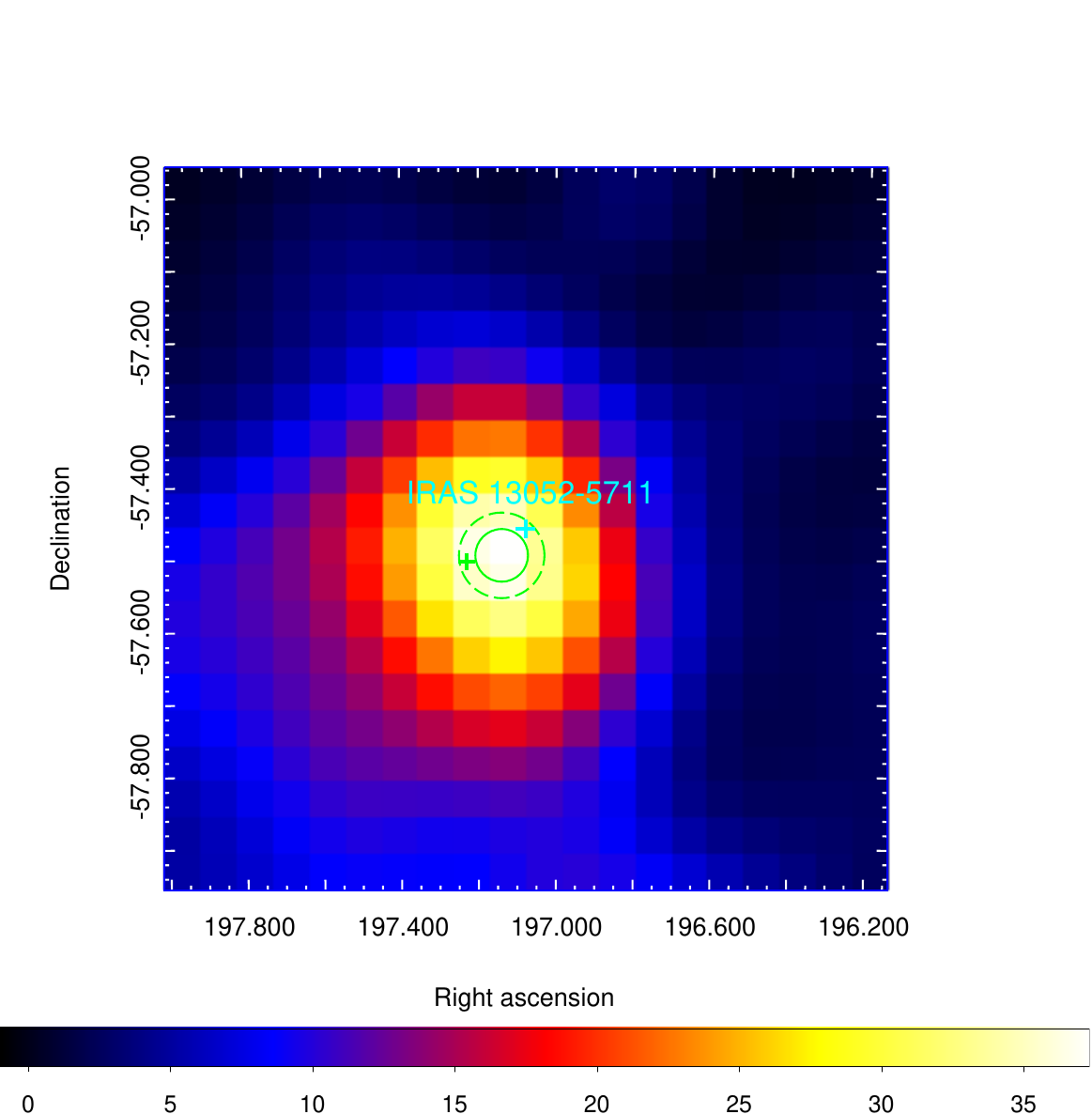}
\caption{  
TS maps of 0.5-500 GeV with 0$^{\circ}$.05 pixel size centered at the position of IRAS 13052-5711, marked with a cyan cross. A Gaussian function  is selected to smooth these TS maps with a kernel radius of 0$^{\circ}$.3. 
Left panel: 5$^{\circ}$.0 $\times$ 5$^{\circ}$.0 of TS map where IRAS 13052-5711 is not deducted.
Middle panel: 5$^{\circ}$.0 $\times$ 5$^{\circ}$.0 of TS map where IRAS 13052-5711 is deducted.
Right panel: 1$^{\circ}$.0 $\times$ 1$^{\circ}$.0 of TS map where the two green circles represent the 1$\sigma$ and 2$\sigma$ error circles of the best-fit location and the green cross represents the position of 4FGL J1308.9-5730.
}  
    \label{Fig1}
\end{figure}

 First, we investigated the location of IRAS 13052-5711 is (R.A., decl. =
197$^{\circ}$.078, -57$^{\circ}$.458) utilizing SIMBAD\footnote{http://simbad.u-strasbg.fr/simbad/}. 
We then investigated the $\gamma$-ray emission at the location of IRAS 13052-5711 in the 4FGL and found that the distance between IRAS 13052-5711 and 4FGL J1308.9-5730 is small, with a value of 0$^{\circ}$.094.
 Using the \textbf{gttsmap} command, we generated a  5$^{\circ}$.0 $\times$ 5$^{\circ}$.0  test statistic (TS) map of IRAS 13052-5711 in the 0.5-500 GeV band, which was selected to avoid an excessively large point spread function, as shown in the left panel of Figure \ref{Fig1}. 
 We found significant $\gamma$-ray  radiation with TS value=43.94 and 6.63$\sigma$ in the center and no significant $\gamma$-ray residual radiation in the surrounding region.

To check whether the $\gamma$-ray radiation originates from IRAS 13052-5711, we used the SIMBAD position of IRAS 13052-5711 to replace the position of 4FGL J1308.9-5730 
and used the Fermitools to deduct the $\gamma$-ray radiation at this position.  We found that the $\gamma$-ray emission at this location can be completely subtracted, as shown in the middle panel of Figure \ref{Fig1}, which indicates that this $\gamma$-ray emission likely originates from IRAS 13052-5711.

Subsequently, the \textbf{gtfindsrc} command was used to generate the best-fit position of the target object (R.A., decl. = 197$^{\circ}$.150, -57$^{\circ}$.495), and its 1$\sigma$ and 2$\sigma$ error circles were 0$^{\circ}$.036 and 0$^{\circ}$.059, respectively. 
We found that the SIMBAD position of IRAS 13052-5711 was within a 2$\sigma$ error circle, indicating that they are related in spatial position. 
We used the best-fit position to replace the position of 4FGL J1308.9-5730 in all subsequent analyses. 
Other possible $\gamma$-ray candidates  were also investigated using SIMBAD, NASA/IPAC Extragalactic Database (NED)\footnote{https://ned.ipac.caltech.edu/} and Aladin\footnote{http://aladin.cds.unistra.fr/aladin.gml}. 
 However, except for IRAS 13052-5711, we did not find other possible $\gamma$-ray candidates, like AGNs, $\gamma$-ray bursts, supernova remnants, pulsars or other galaxies.
Therefore, we deduced that IRAS 13052-5711 remains the most likely candidate for the detected $\gamma$-ray emission at this location.

\setlength{\belowcaptionskip}{0.3cm}

\subsection{\rm $\gamma$-Ray Spatial Distribution Analysis}
We tested the $\gamma$-ray spatial distribution of IRAS 13052-5711 in the 1-500 GeV energy band, which was selected to avoid a large point spread function in the low energy range. 
We chose two-dimensional Gaussian and uniform disk templates to separately test the target source. The independent variables $\sigma$ and radius of the two spatial templates ranged from 0$^{\circ}$.1 to 1$^{\circ}$.0, and the increment was set to 0$^{\circ}$.05. 
We found that the TS value of the target source showed a gradually decreasing trend with the increase of $\sigma$ and radius.
 When the values of $\sigma$ and radius were 0$^{\circ}$.1, the TS value of the target source reached its maximum. 
 We then calculated $\rm TS_{ext} =2log( L_{ext}/L_{ps})$ \citep{Lande2012} and found that the $\rm TS_{ext} \approx8$  for the two models. Because the $\rm TS_{ext}<$16,   
the $\gamma$-ray emission of IRAS 13052-5711 had no significant extension  features \citep{Acero2016}.

\subsection{\rm Spectral Analysis} \label{sec:data-results}

\begin{figure}[!h]
  \centering
  \includegraphics[width=110mm]{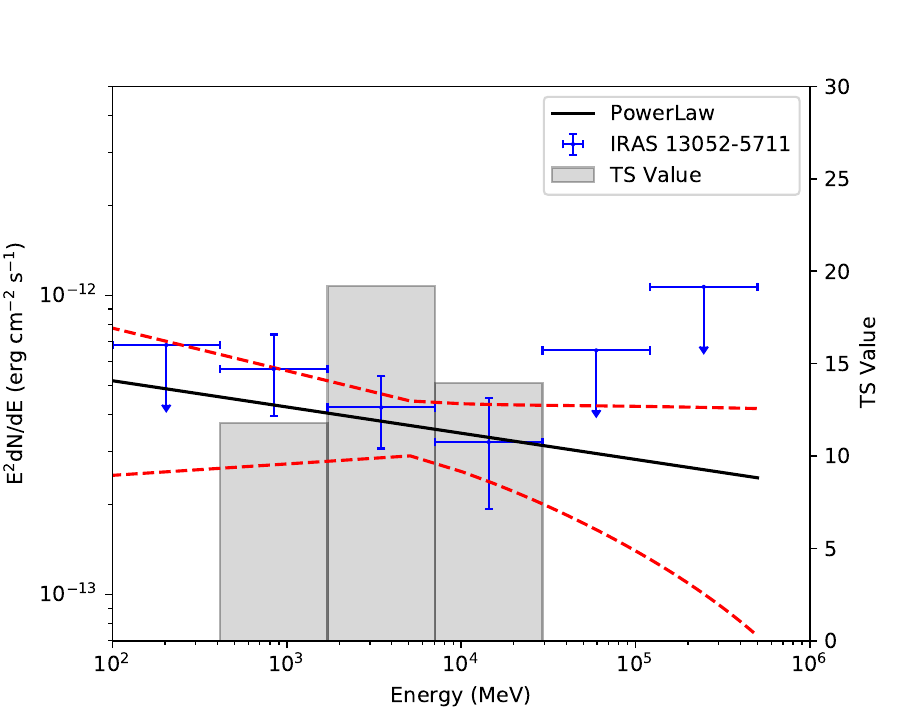}
 \flushleft
\caption{ 
SED of IRAS 13052-5711 in the 0.1-500 GeV energy band. 
The global best-fit power-law spectrum and its 1$\sigma$ statistic error are plotted with 
the black solid and red dashed lines, respectively. 
Blue data points are from the Fermi-LAT data. The best-fit result for the global fit is represented by a solid black  line. The grey shaded area represents the TS value of each energy bin. 
}
\label{Fig2}
\end{figure}

In this analysis, three frequently used spectral models were employed: powerLaw (PL), LogParabola (LOG), and PLSuperExpCutoff (PLEC)\footnote{https://fermi.gsfc.nasa.gov/ssc/data/analysis/scitools/source{\_}models.html}. 
The binned likelihood analysis method  was used to globally fit the photons of IRAS 13052-5711 in the energy range of 0.1-500 GeV. 
 We found that the $\rm TS_{curve}$\footnote{$\rm TS_{curve}$ is defined as 2(log $L$(curved spectrum) - log $L$(powerlaw)), and $\rm TS_{curve}>$16 suggests a spectrum has a significant curve referring to  \citet{Nolan2012}.} values of LOG and PLEC were approximately equal to 2. 
The fitting results showed no significant curvature variability in the spectrum of the target source. 
 Therefore, the PL was selected as the best spectral model for all subsequent analyses. 
 The best-fit spectral index of the PL  was 2.09$\pm$0.13, flux of photons was (2.98$\pm$1.04) $\times\ 10^{-9}$ cm$^{-2}$ s$^{-1}$, and TS value was 42.89 with a significance level of 6.55$\sigma$. 
The spectral energy distribution (SED) of IRAS 13052-5711 was then divided into six equal logarithmic energy bins, and each energy bin was fitted independently by using the binned likelihood analysis method. 
 For an energy bin of TS$<$4, the upper limit was set at a 95\% confidence level, and the relevant results are shown in Figure  \ref{Fig2}.

\subsection{\rm Variability Analysis} \label{sec:data-results}

\begin{figure}[!h]
\centering
 \includegraphics[width=\textwidth, angle=0,width=140mm,height=70mm]{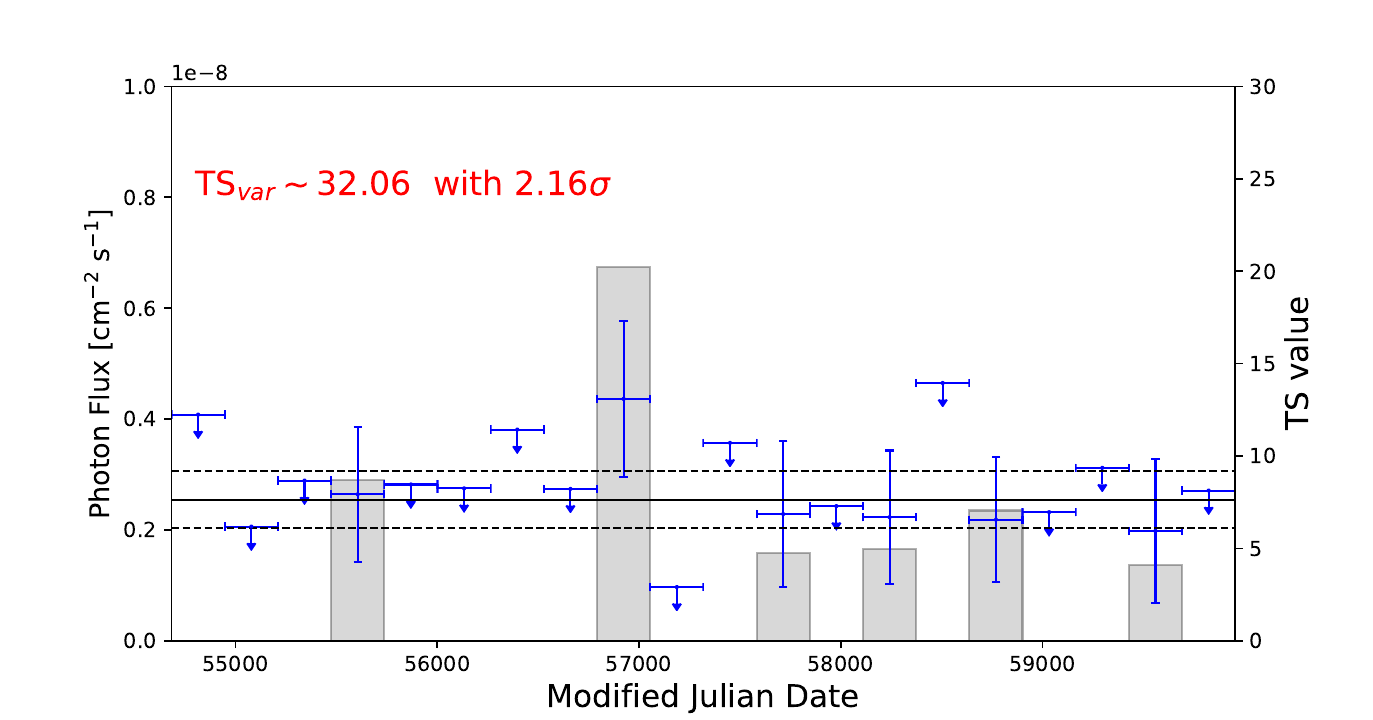} %
 \caption{Panel comprises the LC over 14.4 years with 20 time bins for IRAS 13052-5711 in the 0.2-500 GeV energy band. The gray-shaded regions show the TS values of each bin.  The solid black line represents the best-fit result of a constant-flux model, with two black dashed lines representing its 1$\sigma$ uncertainties.
}
 \label{Fig3}
\end{figure}

To test whether  a significant variability exists in the photon flux of IRAS 13052-5711, we analyzed the LC of IRAS 13052-5711 for 14.4 years. 
To avoid pollution caused by a large point spread function in the low-energy band, we deducted the data below 200 MeV and used  Fermitools to generate the LC between 200 MeV and 500 GeV, as shown in Figure \ref{Fig3}. 
 For the LC with 20 time bins, the variability index, $\rm TS_{\rm var} \geq$ 36.19, of the target source is determined to be a variable source with a 99\% confidence level  \citep{Nolan2012}.  
We calculated $\rm TS_{\rm var}$ to be 32.06, with a variability significance level of 2.16$\sigma$. This result  indicated that IRAS 13052-5711 has no significant flux variability.  
To further verify this result, a  constant-flux model was selected to fit the data points with TS$>$4 in the LC. 
Finally, we found that the reduced $\chi^2$ (=$\chi^2/\rm N_{dof}$) of a  constant-flux model was approximately 0. This result further demonstrated that IRAS 13052-5711 is a stable source without significant variability.

\subsection{\rm Luminosity correlation analysis} \label{sec:data-results}

\citet{Thompson2007} predicted a linear relationship between the total infrared luminosity and the ${\gamma}$-ray luminosity  for dense starbursts.  
\citet{Ackermann2012} established an empirical relation for local group galaxies and SFGs between the total infrared (8-1000 $\mu$m) and the ${\gamma}$-ray luminosity  with the detection of high-energy ${\gamma}$-ray emission of star-forming systems.  
The scalar relationship between $L_{\rm 0.1-500\ GeV}$ and $L_{\rm 0.8-1000\ \mu m}$  can be expressed as follows:

\begin{equation}
{\rm log}\left( \frac{L_{0.1-500 \rm \ GeV} }{\rm erg\ s^{-1}} \right)= {\alpha}\ {\rm log}\left(\frac{L_{8-1000 \rm \ \mu m} }{10^{10}L_{\bigodot}}\right) + \beta,
 \label{eq1}
\end{equation}

Using the BinnedAnalysis function from the Fermitools\footnote{https://fermi.gsfc.nasa.gov/ssc/data/analysis/scitools/python-usage-notes.html}, we calculated the luminosity of IRAS 13052-5711 to be (3.28$\pm$0.67)$\rm \times 10^{42}\ erg\ s^{-1}$  ($d$=93.8 Mpc, from  \citet{Crook2007}) in the 0.1-500 GeV energy band.  
Furthermore, the $L\rm _{\rm 0.1-500\ GeV}$ of other known 11 SFGs and SBGs  were recalculated using the same period and method as those for IRAS 13052-5711,   and the results are shown in Table \ref{Table1}. 
We compared the $L\rm _{\rm 0.1-500\ GeV}$ values of each SFG and SBG with those of  \citet{Guo2019} and \citet{Ajello2020} and found no statistical difference.

\begin{figure}[!h]
\centering
 \includegraphics[width=\textwidth, angle=0,width=75mm,height=60mm]{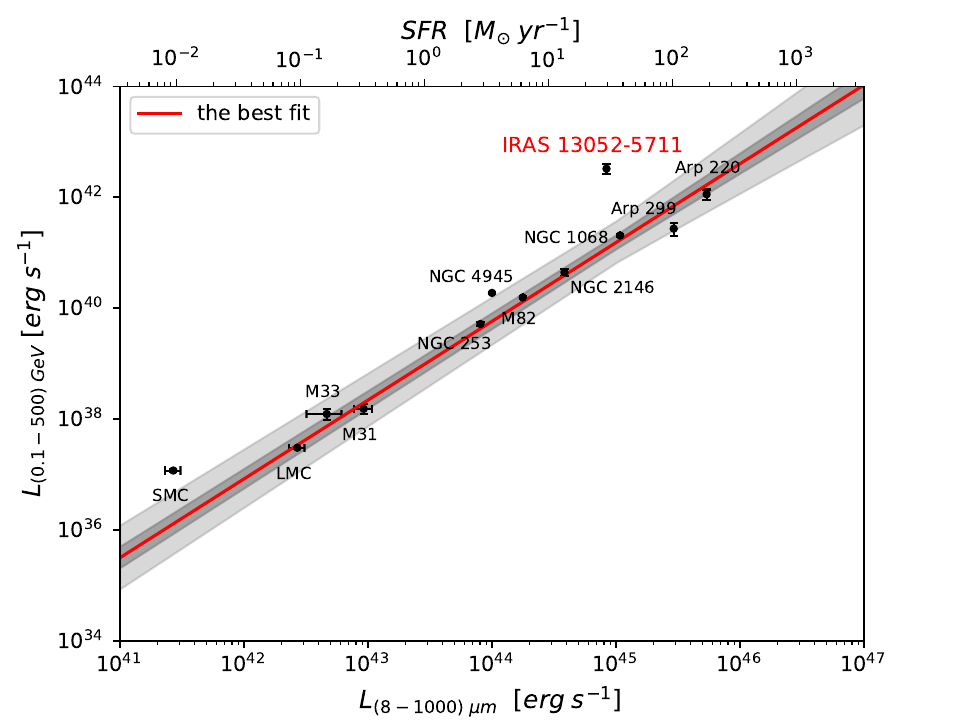}
 \includegraphics[width=\textwidth, angle=0,width=75mm,height=60mm]{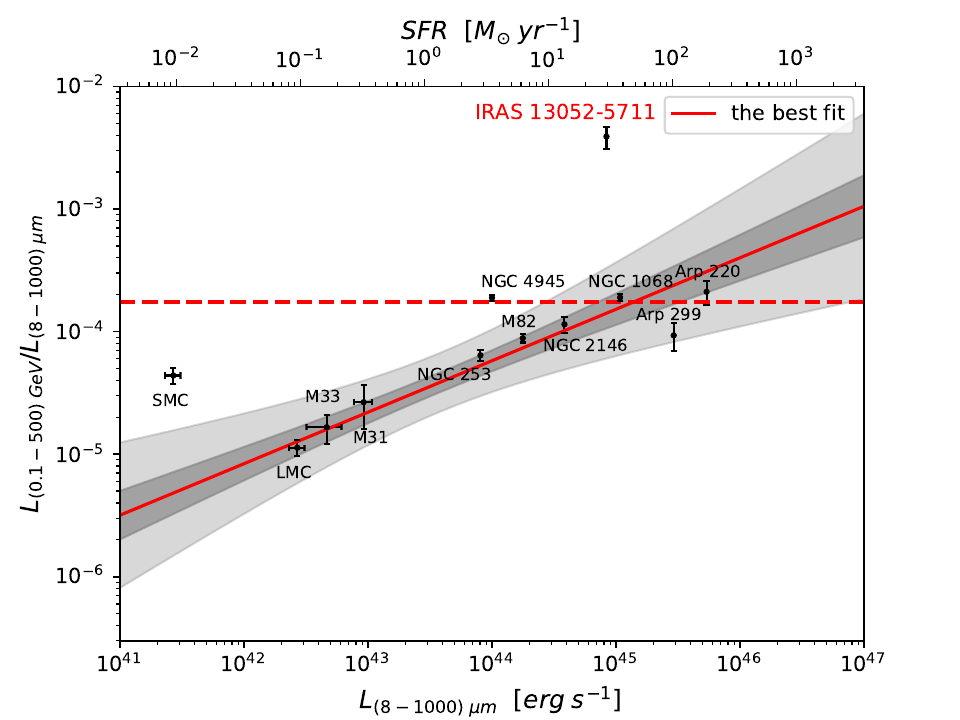}
 \caption{Left panel:  ratio between $L_{\rm 0.1-500\ GeV}$ and $L_{\rm 8-1000\ \mu m}$. Right panel: ratio between $L_{\rm 0.1-500\ GeV}$/$L_{\rm 8-1000\ \mu m}$ and $L_{\rm 8-1000\ \mu m}$. The best-fit relation is plotted as the solid red line;  1$\sigma$ uncertainty is shown as the dark grey shaded region, and its 3$\sigma$ uncertainty is shown as the light grey shaded region. The red dashed line indicates the proton calorimetric limit in the right panel.
}
 \label{Fig4}
\end{figure}

Subsequently, formula (\ref{eq1}) and the curve-fit() function was used in the scipy software package \citep{Virtanen2020} to fit the $L\rm _{\rm 0.1-500\ GeV}$ data of  11 known SFGs and SBGs to obtain  $\alpha=1.41\pm0.07$ and $\beta=39.17\pm0.08$. 
The related fitted results are shown in the left panel of Figure \ref{Fig4}. 
The luminosity of IRAS 13052-5711 is marked in the left panel of Figure \ref{Fig4}; it deviates from the correlation in Equation (\ref{eq1}). 
To visually verify whether IRAS 13052-5711 exceeds the limit of the proton calorimeter, we performed a correlation analysis between $L_{\rm 0.1-500\ GeV}$/$L_{\rm 8-1000\ \mu m}$ and $L_{\rm 8-1000\ \mu m}$, similar to \citet{Ackermann2012} and \citet{Guo2019}, then we found it exceeds this limit, as shown in the right panel of Figure \ref{Fig4}.

In addition, 
the correlation between $L_{\rm 1.4\ GHz}$ and $L_{\gamma}$ is important \citep{Yun2001}. 
However, investigating the current literature and the NED database looking for 1.4 GHz luminosity of IRAS 13052-5711, we haven't found any measurement. 
Therefore, we did not analyze the correlation between $L\rm _{\gamma}$ and $L\rm _{1.4\ GHz}$ in this study.

\begin{table*}[!h]
\begin{center}
\caption{Parameters of the SBGs and SFGs}
\setlength{\tabcolsep}{0.8mm}{
\begin{tabular}{ccccccc}
    \hline\noalign{\smallskip}
     Galaxy & Distance     & $L_{\rm 8-1000\ \mu m}$   & SFR      & $L_{\rm 0.1-500\ GeV}$  & $L_{\rm 0.1-500\ GeV}/ L_{\rm 8-1000\ \mu m}$ & \\
     Name   & D(Mpc)      &  erg s$^{-1}$  & $\rm M_{\odot}$ ${\rm yr}^{-1}$  & erg s$^{-1}$      &  &   \\
  \hline\noalign{\smallskip}
    SMC       &      0.06 & (2.69$\pm$0.38)e+41   & (9.43$\pm$1.33)e-3 &  (1.18$\pm$0.07)e+37  &  (4.38$\pm$0.67)e-5    \\
    LMC       &      0.05 &  (2.69$\pm$0.38)e+42  & (9.43$\pm$1.33)e-2   &  (3.03$\pm$0.11)e+37  & (1.13$\pm$0.17)e-5 \\
    M31       &    0.78   &   (9.23$\pm$1.54)e+42 & (3.24$\pm$0.54)e-1  &  (1.53$\pm$0.31)e+38  & (2.65$\pm$1.04)e-5 \\
    M33  &   0.85   &  (4.65$\pm$1.46)e+42        & (1.63$\pm$0.51)e-1  &  (1.23$\pm$0.29)e+38   &  (1.66$\pm$ 0.44)e-5 \\
    NGC 253   &   2.5    &  8.07e+43              &  2.83  &  (5.17$\pm$0.49)e+39     &    (6.41$\pm$0.61)e-5    \\
    M82       &     3.4  &  1.77e+44         &  6.21    &  (1.56$\pm$0.12)e+40     &  (8.82$\pm$0.68)e-5 \\
    NGC 4945  &    3.7   &   1.00e+44        &  3.51    &  (1.88$\pm$0.10)e+40     &   (1.88$\pm$0.10)e-4  \\
    NGC 2146  &    15.2  &  3.84e+44        &  1.35e+1     &  (4.40$\pm$0.67)e+40    &   (1.15$\pm$0.17)e-4   \\
    NGC 1068  &  16.7     &  1.08e+45       &  3.79e+1     &  (2.05$\pm$0.14)e+41     &    (1.89$\pm$0.13)e-4   \\
    Arp 299   &    44.0  &  2.92e+45       &  1.02e+2       &  (2.71$\pm$0.69)e+41    &    (9.30$\pm$2.37)e-5   \\
    Arp 220   &    74.7   &  5.38e+45         &  1.89e+2  &  (1.13$\pm$0.25)e+42    &   (2.11$\pm$0.46)e-4    \\
IRAS 13052-5711 &   93.8   &  8.38e+44       &  1.48e+1   &  (3.28$\pm$0.67)e+42     &   (3.90$\pm$0.80)e-3     \\
  \noalign{\smallskip}\hline

\end{tabular}}
\end{center}
Note: 
\textcolor{black}{
Distance of IRAS 13052-5711 is obtained from \citet{Crook2007}; that of Arp 299 is from \citet{Pereira2010}; distances of other sources are from \citet{Ackermann2012} and references therein. 
$L_{8-1000\mu m}$ of IRAS 13052-5711 and Arp 299 are obtained from \citet{Sanders2003}; 
those of other sources are from \citet{Ackermann2012} and references therein. The $\gamma$-ray luminosities in 0.1-500 GeV are calculated by using the same period as IRAS 13052-5711 in this work.}
  \label{Table1}
\end{table*}

\section{DISCUSSION} \label{sec:data-results}   

\subsection{\rm Authentication of source}
Through cross-identification of spatial locations, we excluded other common $\gamma$-ray candidates, and found that IRAS 13052-5711 and the $\gamma$-ray radiation at this position are interrelated. 
Based on this, we analyzed the spectral characteristics. 
We found that its spectrum had no significant curvature variability with $\rm TS_{curve}\approx2$ and could be  fitted by a simple PL spectral function. In addition, its PL spectral index  was approximately 2.1, which is close to 2.2 of the SBGs reported thus far. 
Moreover, we explored the variability characteristic of the LC over 14.4  years. 
We found that the LC has no significant variability, with a variability significance level of 2.16$\sigma$, and can be well fitted by employing a constant-flux model with a reduced $\chi^2\sim$0. 
This stable characteristic of the photon flux is consistent with those of the SBGs found to date. 
Additionally, the SFR is regarded as an important evaluation indicator for identifying SBG, which is calculated using the formula (\ref{eq2}) from  \citet{Kennicutt1998}. The reported relation is the following:

\begin{equation}
 \frac{\rm SFR} {\rm M_{\odot}\ {\rm yr}^{-1}} = 1.7\epsilon \times 10^{-10} \frac{L_{8-1000 \rm \ \mu m} }{L_{\bigodot}},
 \label{eq2}
\end{equation}
where $\epsilon$ is a function dependent on the initial mass and its value is taken as 0.79 \citep{Ackermann2012}; the solar bolometric luminosity $L_{\bigodot}=3.83\times 10^{33}\ \rm erg\ s^{-1}$  \citep{Sanders2003}. 
We obtained SFR $\approx$ 29.38 $\rm M_{\odot}$ $\rm yr^{-1}$ as per Equation (\ref{eq2}); this value is within the range of the SFR ($>$2 $\rm M_{\odot}\ yr^{-1}$) calculated by $L_{\rm 8-1000\ \mu m}$ for the currently discovered SBG candidates \citep{Ackermann2012}. 
Based on the foregoing discussion, we believe that IRAS 13052-5711 is likely  an SBG.

\subsection{\rm The $\gamma$-ray origin of IRAS 13052-5711}

We analyzed whether IRAS 13052-5711 conformed to the well-known correlation between the $\gamma$-ray luminosity and total infrared luminosity.
 By using 14.4 years of  Fermi-LAT data, we reanalyzed the luminosity correlation from 11 SBGs and SFGs and found that values of 14.4 years of GeV luminosity and the best-fit result of $\alpha= 1.41\pm0.07$ and $\beta = 39.17\pm0.08$ are approximately consistent with those obtained by \citet{Guo2019} and \citet{Ajello2020}. 
Subsequently,  IRAS 13052-5711 was found to be
 beyond the relationship between $L_{\rm 0.1-500\ GeV}$ and $L_{\rm 8-1000\ \mu m}$ and to lie above the proton calorimeter limit. For the current luminosity analysis results, we considered the following two scenarios:

Scenario 1: The $\gamma$-ray radiation is produced by the AGN hosted inside the SBG.  
When a galaxy harbors an active nucleus, a supermassive black hole powers $\gamma$-ray emission. 
The radiation provided by the AGN and the inner star-forming region exceeds the limit of the proton calorimeter \citep{Wang2018}. 
Currently, exploring the time-varying characteristic of LC as a common means can help us identify whether there is a contribution from the AGN. 
This method has been used in previous studies, such as for Circinus \citep{Guo2019}, NGC 3424, UGC 11041 \citep{Peng2019}, and Arp 299 \citep{Xi2020a}. 
However, by investigating the time-varying characteristic of the 14.4 years of the LC of IRAS 13052-5711, we did not find any significant variability. 
 Hence, our analysis does not support the AGN scenario.

Scenario 2: IRAS 13052-5711 might be a proton calorimeter, even if the $\gamma$-ray luminosity is above the Kennicutt relation as shown in Figure \ref{Fig4}. This may occur, in fact, under the assumption that SNe inside IRAS 13052-5711 have a high particle acceleration efficiency $\eta$ than what is commonly assumed for Milky Way SNR. According to Equation (3) derived from \citet{Guo2019}, a larger value of $\eta$ increases the ratio $L_{\rm 0.1-500\  GeV}/L_{\rm 8-1000\ \mu m}$ causing a deviation from the empirical relation given in formula (\ref{eq1}).

Previous studies have shown that Arp 220 is a probably hadronic calorimeter and that the internal proton and proton interaction (PP) dominates its internal $\gamma$-ray radiation \citep{Wang2018, Peng2016, Torres2004, Lacki2013, Yoast2015}.
   \citet{Wang2018} considered Arp 220 to harbor a large number of SNe that  could systematically  and more efficiently accelerate the internal CRs than in the  Milky Way under the assumption of a proton calorimeter.  
SNe inside the Milky Way are usually assumed to have an $\eta$ of 3-10\% \citep{Strong2010}. \citet{Wang2018} assumed that the SNe in the proton calorimeter have a higher $\eta\sim$30\%  than the one in the Milky Way, and the proton injection spectral index is between 2.0 and 2.4. 
 However, \citet{Wang2018} found that Arp 220 was still above the inferred proton calorimeter limit, suggesting that the $\eta$ of protons in CRs per SNR inside Arp 220 could be higher than those of previously discovered SBGs, excluding the contribution of the AGN. 
As the LC of IRAS 13052-5711 does not exhibit  significant variability and the $\gamma$-ray luminosity and distance of  IRAS 13052-5711 are close to those of Arp 220 \citep{Peng2016, Xiang2021a}, the scenario in which IRAS 13052-5711 is a proton calorimeter cannot be ruled out.

\begin{figure}[!h]
\centering
 \includegraphics[width=\textwidth, angle=0,width=80mm,height=60mm]{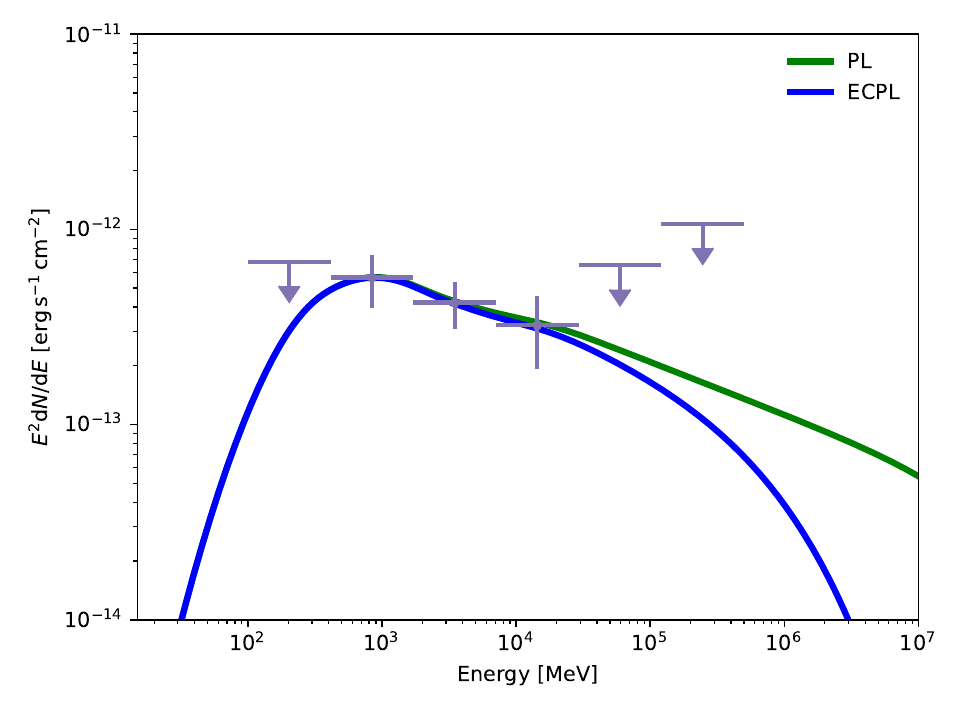}
 \caption{The solid green and blue lines represent the best-fit results of the PL and ECPL models, respectively.
}
 \label{Fig5}
\end{figure}

Based on the assumption of proton calorimeter, we selected a hadronic model to simulate the spectrum of IRAS 13052-5711. Two types of particle energy distribution models were selected, and their expressions are as follows:

Model 1: a power law function (PL):
\begin{equation}
 N(E) = N_{0}\left (\frac{E}{E_{0}}\right )^{-\alpha},
\end{equation}

Model 2: a power law with an exponential cutoff function (ECPL):
\begin{equation}
N(E) = N_{0}\left (\frac{E}{E_{0}}\right )^{-\alpha}exp\left (-\frac{E}{E_{\rm cutoff}}\right),
\end{equation}
where $N_{\rm 0}$ is the amplitude, $E$ is the particle  energy, $\alpha$ is the spectral index, and $E_{\rm cutoff}$ is the cut-off energy, and $E_{\rm 0}$ is adopted to be 1 TeV.

In the case of PL, \citet{Abdalla2018} found that NGC 253 exhibited a hard spectral signature with a spectral index$\sim$2.1 of Fermi-LAT.  They used the PL particle energy spectrum to fit the observed data. As IRAS 13052-5711 also presents a hard spectrum with the
 same spectral index of $\sim$2.1, we adopted the PL model for this analysis.

We also decided to test the ECPL looking for the possible presence of a cutoff. The presence of a cutoff is indeed expected in the scenario where particles are accelerated at SNRs inside the starburst nucleus, however, if the process is similar to the Milky Way, we do expect a proton cutoff at energies $>$ 100 TeV, hence for the $\gamma$-ray spectrum the cutoff energy should be located at E$>\sim$10 TeV. Actually, given the larger number of SNR in SBGs, it is probable that the maximum energy of accelerated protons is even larger than the one in the Milky Way \citep[see][]{Peretti2019}. Hence the possible presence of a cutoff in the Fermi-LAT energy range would imply a physical mechanism rather different from the standard SNR acceleration.

A spectral analysis tool named Naima \citep[][and references therein]{Zabalza2015}, was used to perform this analysis. Here, the PP differential cross-section of PYTHIA 8 \citep{Kafexhiu2014} was selected to fit the spectrum of IRAS 13052-5711. 
For the absorption of extragalactic background light (EBL), here we considered the EBL model from \citet{Dominguez2011} and used redshift z=0.02 \citep{Strauss1992} of IRAS 13052-5711 in all the following analyses. 
Considering that the distance of IRAS 13052-5711, the spectral index, GeV photon flux, and gamma-ray luminosity are close to those of Arp 220 \citep{Peng2016, Xiang2021a}, we assumed that they have a close interstellar medium density $n_{\rm ISM}$=3500 $\rm cm^{-3}$,  referring to  \citet{Peretti2019}. 
The PL and ECPL were selected to fit the Fermi-LAT observed data points in Figure \ref{Fig2}. 
The Bayesian information criterion (BIC) was used to determine the goodness of fit between the two models \citep{Schwarz1978, Ambrogi2019}. 
The related BIC formula is $\rm log(n)k - 2log(L)$, where n represents the number of the observed data points, k represents the number of model parameters, and L is the maximum likelihood value. The model with the lowest BIC value is preferred. The relevant parameters obtained are listed in Table \ref{Table2}.

\begin{table*}[!h]
\caption{The Best-fit Parameters of PL and ECPL Models}
\begin{center}
\begin{tabular}{cccccccc}
    \hline\noalign{\smallskip}
  Model  & $\alpha$    & $E_{\rm cutoff}$   &  $W_{\rm p}$ &   Log(L) & BIC & $\chi^{2}/\rm N_{dof}$ \\
                   &       & TeV   & erg   & & \\
  \hline\noalign{\smallskip}
  \multirow{1}{*}{PL}   

  & 2.22$_{-0.18}^{+0.15}$  & --- &  9.87$_{-0.61}^{+1.81}\times 10^{54}$ & -0.00023 & 5.376 & $\frac{0.00023*2}{6-4}=0.00023$ \\
      \noalign{\smallskip}\hline
  \multirow{1}{*}{ECPL}  
  
   & 2.26$\pm$0.15  & 10.92$_{-2.28}^{+2.45}$ &  9.73$_{-3.78}^{+4.36}\times 10^{54}$ & -0.00079 & 7.169 & $\frac{0.00079*2}{6-4}=0.00079$ \\
   \hline\noalign{\smallskip}
\end{tabular}
\end{center}
\label{Table2}
Note: Hadronic energy budget, $W_{\rm p}$, is calculated at above  290 MeV \citep{Abdalla2018}. Log(L) represents the maximum of the log(likelihood). Three upper limits of the SED are included in the fitting \citep[e.g.,][]{Abdalla2018} and $\rm \chi^{2}=-2log(L)$ \citep{Zabalza2015}. 
\end{table*} 

By conducting numerical simulation, we found that 
the values of reduced $\chi^{2}$ of PL and ECPL models are approximately 0. By comparing the BIC values obtained from the fitting of two models, we found no significant difference, as shown in Table \ref{Table2}. 
These results show that both PL and ECPL can both well describe the $\gamma$-ray emission from IRAS 13052-5711, as shown in Figure \ref{Fig5}. However, the cutoff value of $\sim$ 11 TeV should only be considered as a lower limit for a possible cutoff, given that the data does not show any clear departure from a simple power-law.

In addition, we find that the proton slope $\alpha$ for the PL model is approximately 2.22, which is in close agreement with that of NGC 253 with $\alpha\sim$2.2 from the work of \citet{Abdalla2018}. Such a hard particle spectrum is consistent with the proton calorimetric scenario in which the equilibrium spectrum inside the starburst nucleus is close to the spectrum injected by SNRs. In fact, the spectral slope of CRs injected by SNRs in the Milky Way is between 2.2 and 2.3 in fully agreement with our result of $\sim$2.2. The slope inferred for IRAS 13052-5711 is also similar to other SBGs like NGC 253 and M 82, while the other famous example, Arp 220, shows a steeper spectrum (2.45 - see \citet{Peretti2019}). The latter case, Arp 220, has parameters very similar to IRAS 13052-5711, hence we compare the proton energy content of the two SBGs: from Table 4 of \citet{Peretti2019} the energy density of the protons is $U_{\rm p}$ = 1324 eV cm$^{-3}$ and the radius is $R$ = 250 pc; hence, $W_{\rm p} =\frac{4}{3}\pi R^{3} U_{\rm p}\approx$4e54 erg. Such a value is only twice smaller that the value we estimated for IRAS 13052-571 (see Table \ref{Table2}), which reinforces the idea that IRAS 13052-5711 could be a proton calorimeter.  
Following \citet{Wang2018}, we have tested the scenario with an acceleration efficiency as large as $\eta$= 30\%, finding that such a value is not enough to explain the observed value of $L_{\rm 0.1-500\ GeV}$/$L_{\rm 8-1000\ \mu m}$ in IRAS 13052-5711. This is in tension with results from the theory of particle acceleration at SNRs, where acceleration efficiency larger than 30\% is difficult to achieve. Possible explanations of such a discrepancy may be: i) the star formation rate is larger than what estimated in this work or ii) there is an additional source of particle acceleration beyond SNRs. One possibility for ii) is the acceleration at the termination shock of stellar winds [see e.g. \citet{Bykov2020}; \citet{Morlino2021}].
Further data below 400 MeV, that may be collected in the future, will help to confirm the hadronic origin of the $\gamma$-ray emission.

Finally, we considered the inverse Compton scattering (IC) and bremsstrahlung radiation (Brem) from relative electronics to interpret the  GeV SED of IRAS 13052-5711. 
For Brem, we took ionized gas density $n_{\rm ion}=87.5\ \rm cm^{-3}$, referring to \citet{Peretti2019}. For IC, the cosmic microwave background radiation field was considered. Through simulation with PL and ECPL, we found that IC and Brem can effectively explain the GeV SED of IRAS 13052-5711 with ${\chi}^2/N_{dof}\sim 0$. The energy contents of leptons are $W_{\rm e,Brem} \sim 4e55 $ erg and $W_{\rm e,IC}\sim 4e60 $ erg.

Now, let us assume that electrons are produced by acceleration in SNRs. The maximum energy that can be accumulated in accelerated electrons is given by
\begin{equation}
W_{\rm e} = \xi_{\rm e}  E_{\rm SN}  \Gamma_{\rm SN}  \tau_{\rm loss}
\end{equation}
Where $\xi_{\rm e}$ is conversion efficiency of SN energy into electrons ($<$1 but usually $\sim$1e-3 for SNR in the Milky Way),
 $E_{\rm SN}$ = 1e51 erg is the average SN explosion energy, 
 $\Gamma_{\rm SN}$ is the SN explosion rate that can be easily estimated from the SFR provided in the paper just below Eq.(\ref{eq2}) (SFR=29.38 $M_{\odot}$/yr)). In fact assuming a Salpeter initial mass function \citep{Salpeter1955}, the rate of SN can be estimated as the rate of production of stars with mass $>$ $M_{\rm c}$ = 8$M_{\odot}$, hence:
 \begin{equation}
 \Gamma_{\rm SN} = SFR * \int_{M_{\rm c}}^{M_{\rm max}} m^{-2.35} dm / ( \int_{M_{\rm min}}^{M_{\rm max}} m*m^{-2.35} dm )
 \end{equation}
 Where $M_{\rm min}$ and $M_{\rm max}$ are the minimum and maximum stellar mass, respectively assumed to be 0.08 and 150$M_{\odot}$.
Finally $\tau_{\rm loss}$ is the loss time for electrons inside the nucleus. This should be properly calculated in a self-consistent model, however, given the already mentioned similarity with Arp 220, we will use the maximum possible value estimated in \citet{Peretti2019} (see their figure 1 top panel), namely $\tau_{\rm loss}\sim10^5$ yr. Using those numbers we get $W_{\rm e} = \xi_{\rm e} *$  2e55 erg. Such a value is a very optimistic upper limit, using more realistic value of $\tau_{\rm loss}\sim 10^3$ yr and $\xi_{\rm e}$= 1e-2, we got $W_{\rm e}\sim$ 2e51 erg. Hence leptonic emission is ruled out in our opinion.

\section{\rm Summary}

Based on the foregoing analyses, we  report a likely SBG, IRAS 13052-5711, which is the most distant SBG candidate  identified thus far.  
We found that its spectrum in the 0.1-500 GeV band is well fitted by a PL model and  shows no significant curvature variability with $\rm TS_{curve}\approx 2$. The photon flux is (2.98$\pm$1.04)$\times 10^{-9}$ $\rm cm^{-2}\ s^{-1}$ and TS value is 42.89 with 6.55$\sigma$. 
 Its spectral slope of $\sim$2.2 is fully consistent with other SBGs emitting in the $\gamma$-ray band known so far.
 In addition,  we found that the LC over 14.4 years has no significant variability with $\rm TS_{var}\approx$32.06, the variability significance level is $\sim$2.16$\sigma$, and the LC can be well fitted by applying a  constant-flux model with a reduced $\chi^{2}$ $\sim$ 0.  
The steadiness of the LC, together with the spectral slope, suggests that the $\gamma$-ray emission comes from SNRs in the starburst nucleus, rather than from a possible AGN activity.
  
  We performed a luminosity correlation analysis between $L_{\rm 0.1-500\ GeV}$ and $L_{\rm 8-1000\ \mu m}$ and found that IRAS 13052-5711 has a large  GeV luminosity value  with $L_{\rm 0.1-500\ GeV}=$(3.28$\pm$0.67)$\rm \times 10^{42}\ erg\ s^{-1}$, which is close to that of Arp 220. 
However, it deviates from the relationship of $ L_{\rm \gamma}-L_{\rm IR}$ and exceeds the proton calorimeter limit. 
 By excluding the contribution from the AGN inside IRAS 13052-5711, we believe that IRAS 13052-5711 is likely a proton calorimeter, in which PP dominates its  $\gamma$-ray radiation.

 We considered the case of PP and used models of PL and ECPL to fit the $\gamma$-ray spectrum of IRAS 13052-5711 and found that the two models can well explain the GeV SED of IRAS 13052-5711.
 However, there is no clear evidence for a cutoff in the spectrum, hence we can derive a lower limit for possible cutoff energy in the proton spectrum by $E_{\rm cutoff,min}\sim$ 11 TeV.
In addition, as its proton spectrum index is $\rm \alpha \approx 2.2$, which is consistent with that of the proton calorimeter, its $W_{\rm p}$ is close to that of Arp 220 within the calorimetric limit.
Therefore, a situation in which IRAS 13052-5711 is a proton calorimeter cannot be ruled out. 
In the future, spectral data below 400 MeV should be recorded to test the pion-decay bump from IRAS 13052-5711 by using Fermi-LAT continuous sky survey observations, which will help us further confirm this idea.

\section{Acknowledgments} 
We sincerely thank the reviewer for his/her valuable  recommendations for improving the quality of this paper. 
We also appreciate the support for this work from the Natural Science Foundation Youth Program of Sichuan Province (2023NSFSC1350), the Doctoral Initiation Fund of West China Normal University (22kE040), the Open Fund of Key Laboratory of Astroparticle Physics of Yunnan Province  (2022Zibian3)
, the Sichuan Youth Science and Technology Innovation Research Team (21CXTD0038), the National Key R\&D Program of China under Grant No. 2018YFA0404204, the National Natural Science Foundation of China (NSFC U1931113, U1738211, 11303012).

\end{document}